\documentclass[conference]{IEEEtran}
\IEEEoverridecommandlockouts

\usepackage{cite}

\usepackage[dvipsnames]{xcolor}
\usepackage{graphicx}
\usepackage{multirow}

\usepackage{amsmath,amssymb,amsfonts}
\usepackage{amsthm}
\usepackage{mathrsfs}
\usepackage{nicefrac}

\usepackage{booktabs}
\usepackage{tabularx}

\usepackage{algorithm}
\usepackage{algorithmicx}
\usepackage{algpseudocode}

\usepackage{textcomp}
\usepackage{manyfoot}
\usepackage{listings}
\usepackage{float}

\usepackage[utf8]{inputenc}
\usepackage[T1]{fontenc}

\usepackage{url}
\usepackage{microtype}

\usepackage[compact]{titlesec}
\titlespacing{\section}{0pt}{1.2ex}{0.8ex}
\titlespacing{\subsection}{0pt}{1ex}{0.5ex}

\usepackage{hyperref}

\hypersetup{
    pdftitle={Multimodal Dual-Encoder Retrieval for Automated ICD Coding},
    pdfauthor={Abhinav Bohra and Anuj Bohra}
}

\def\BibTeX{{\rm B\kern-.05em{\sc i\kern-.025em b}\kern-.08em
    T\kern-.1667em\lower.7ex\hbox{E}\kern-.125emX}}

\begin{document}

\title{Multimodal Dual-Encoder Retrieval for Automated ICD Coding}

\author{
\IEEEauthorblockN{Abhinav Bohra}
\IEEEauthorblockA{
\textit{Amazon} \\
Seattle, USA \\
bohraabhinav8@gmail.com \\
ORCID: 0009-0008-8377-0604
}
\and
\IEEEauthorblockN{Anuj Bohra}
\IEEEauthorblockA{
\textit{Rutgers University} \\
New Jersey, USA \\
anujbohra232003@gmail.com \\
ORCID: 0009-0005-0532-0745
}
}

\maketitle 

\begin{abstract}
Accurate International Classification of Diseases (ICD) coding is crucial for large-scale clinical research, documentation, and billing. There are three primary problems with current ICD prediction methods: (1) They are unable to comprehend multimodal patient data because they rely on either structured EHR data or unstructured clinical notes. (2) They also struggle with scalability to a larger amount of ICD codes (9K+ codes in ICD-9), as traditional classifiers need dense output layers and often do not generalize well to long tail rare diseases. (3) They lack transparency for clinical use. To address these challenges, this research proposes a two-stage framework that first retrieves ICD codes using a multimodal dual-encoder retrieval model, where structured and unstructured patient data are integrated through gated fusion. The second stage refines the top-k retrieved candidates with an LLM-based re-ranker that provides ranked codes with clinically relevant explanations. Our experiments show that the proposed approach improves Micro-F1 and Precision over a multimodal dual-fusion classifier baseline. These improvements demonstrate that combining a gated multimodal retrieval system with LLM-based re-ranking is a practical alternative to dense multi-label classification for automated ICD coding.
\end{abstract}
\begin{IEEEkeywords}
Automated ICD coding, Healthcare, Multimodal Fusion, Dual-encoder retrieval, ANN(FAISS), LLM-based re-ranking, Clinical NLP
\end{IEEEkeywords}
\section{Introduction}
Healthcare systems around the world use a standardized coding system to classify and document patient conditions. The International Statistical Classification of Diseases (ICD), created by the World Health Organization (WHO), serves as the global standard for documenting patient conditions, facilitating medical research, and the development of predictive models\cite{bib1} \cite{nelson2024_icd_codes_reliability} \cite{yuan-etal-2022-synonyms} Today, medical codes are still largely assigned manually by medical professionals. This leads to scalability issues due to an increase in multimodal data in Electronic Health Records (EHR), ranging from structured clinical lab reports to unstructured clinical notes captured during a patient's visit. Combined with a poor doctor-to-patient ratio in some developing countries ranging from 1:1456 to 1:10000 \cite{Ugargol2023PHCIndia}, which is lower than the recommended ratio of 1:1000 by WHO \cite{bib3}, this creates a pressing need for automated ICD coding to reduce manual workload and support clinical decision making.

Retrieval systems based on dual-encoder architectures are well known to efficiently match queries with items at scale. This design is widely deployed in recommendation platforms such as YouTube \cite{covington2016_youtube}, because it decouples query computation from item indexing, enabling the model to perform faster candidate generation over millions of items using an approximate nearest neighbor (ANN) search \cite{ANN}. ICD coding faces a similar scalability constraint where each patient admission needs to be compared against a large code catalog, making multi-label scoring expensive as the label space grows. Motivated by this setting, we adopted the dual-encoder retrieval framework to encode patient admissions and ICD code data into a shared embedding space\cite{DBLP:journals/corr/abs-2004-04906} to efficiently output top-k candidate codes. Our methodology integrates two complementary modalities to create a multimodal patient profile: Structured lab reports and unstructured clinical notes are provided to an encoder with gated fusion to combine embeddings from both modalities and form a unified patient embedding. In parallel, the ICD encoder embeds code metadata into the same space. The dual-encoder retrieval model learns to bring clinically appropriate ICD codes close to their corresponding patient representation in the embedding space. The retrieval output consists of top-k candidate codes with similarity scores indicating relative relevance. The subsequent step is a re-ranker to further refine the ICD codes ranking. We use an LLM to rescore the retrieved top-k candidate codes by providing both patient and ICD code data as input. The explanation-aware re-ranking improves the ranking quality by using LLM's semantic understanding and reasoning capabilities, while simultaneously providing a brief natural language explanation to support human review. The primary objective of this work is to assist healthcare professionals in their screening process. Empirical results on benchmark datasets such as MIMIC-III \cite{bib2} demonstrated a consistent improvement in performance relative to existing baselines. The main contribution is a scalable multimodal retrieve-and-rerank approach for ICD coding, where patient admissions and ICD-code metadata are aligned in a shared embedding space and retrieved codes are further refined using an explanation-aware re-ranker.

\vspace{-1mm}
\section{Related Work}\label{sec2}
This section provides the advancements in ICD code prediction. Automated ICD code prediction has been a point of concern in the medical industry due to the high complexity of codes and the high cost of manual coding. 

Automated assignment of ICD codes has been  researched using deep neural architectures over clinical notes as input. Mullenbach et al. (2018) \cite{mullenbach2018_explainable_medical_codes} proposed Convolutional Attention for Multi-Label Classification (CAML), it applies a CNN encoder with label specific attention, where each ICD code is associated with an attention vector to focus on relevant portions of the patient discharge summary. Trained on the MIMIC-III dataset, it achieves a Micro-F1 score of 54\%. However, the architecture appears less attractive for large label spaces (e.g., full ICD-9 with 9k+ labels), due to its classification head with one sigmoid output per code. Vu et al. (2020) extended this work with the Label Attention Model (LAAT) as an improvement over the CAML architecture. LAAT introduces BiLSTM text encoder with a label specific attention mechanism to capture long range dependencies and achieves a higher Micro-F1 score(58\%). However, LAAT still lacks scalability due to the same fundamental limitations of dense output layer with one classifier per ICD code, which scales linearly with the number of labels.

A multimodal framework for automated ICD coding \cite{xu2019multimodal} uses unstructured clinical notes, semi-structured diagnosis descriptions, and structured tabular EHR data. This architecture integrates a Text-CNN classifier for free-text notes, a diagnosis-ranking network using character level CNN + BiLSTM for semi-structured text, and a decision-tree model for structured lab data. These models are combined using an ensemble learning approach that improves classification performance. It achieves a Micro-F1 score of 76.33\%. However, this framework still treats ICD assignment as a multi-label classification problem with separate modality specific models and a shared output layer, which limits scalability to large code spaces. TreeMAN \cite{Treeman} extends multimodal ICD coding by enriching unstructured text data with structured data derived through decision tree models. Structured EHR attributes are first distilled into tree based feature vectors which are then fused with clinical notes embeddings using a hierarchical attention mechanism. This multimodal integration improves the model's ability to capture both clinical context and structured signals, thereby addressing a standard limitation of text-only ICD classifiers. Experiments on benchmark datasets demonstrate that TreeMAN outperforms purely text-based baselines and also demonstrates the value of multimodal data for clinical coding tasks. The method remains a classification model and does not scale well in large retrieval systems with re-ranking components, which limits its scalability to very large code sets. 

A prominent method to address scalability is the multi-stage retrieve-and-re-rank model \cite{wang2024_multistage_retrieve_rerank}, which considers the task as a medical code recommendation problem rather than a conventional multi-label classification. The initial retrieval stage uses hybrid keyword and semantic search (including BM25) to narrow down ICD codes, followed by a re-ranker with contrastive learning to improve the prediction accuracy. On MIMIC-III dataset, it achieves a Micro-F1 of 60\% proving the ability of retrieval and re-ranker models as compared to the traditional classification heads. Niu et al. \cite{niu2023_retrieve_rerank} also proposed a two-stage retrieve-and-re-rank framework. In the first stage, it uses a contrastive dual encoder to embed clinical notes and ICD codes into a shared semantic space. In the second stage, a Transformer based re-ranker takes each patient-code pair as input and refines the ranking. On MIMIC-III dataset, it achieves a Micro-F1 of 59\%. However, both approaches provide clinical notes as input, and do not leverage structured EHR features, and provide limited explicit interpretability. 

The paper by Yan et al. \cite{yan2022survey} provides a survey on automated ICD coding, marking the change from rule based systems to deep learning architectures. The author highlights the large scale of codes, class imbalance, multimodal EHR integration and the need for explainability in new architectures for more clinical relevance.

Across previous studies, text only and multimodal classifiers suffer from scalability limitations due to dense output layers and limited support for long tail codes. Retrieval based and label semantic methods improved scalability and ranking metrics by leveraging ICD code descriptions and contrastive learning, but they are limited to text only and offered limited explainability. Our framework unifies these strengths by learning a multimodal patient representation through gated fusion of unstructured clinical notes and structured EHR data, retrieving codes through a dual-encoder architecture aligned in a shared semantic space, refining predictions with an LLM based re-ranker that also provides natural language justification. This combination directly addresses scalability, multimodal integration, and interpretability. 
\vspace{-1.5mm}
\section{Methodology}\label{sec3}
This section covers the proposed architecture for the multimodal ICD code retrieval, followed by LLM re-ranker to refine the ordering of the retrieved candidates.

\subsection{Training Architecture}

\begin{figure}
    \centering
    \includegraphics[width=1\linewidth]{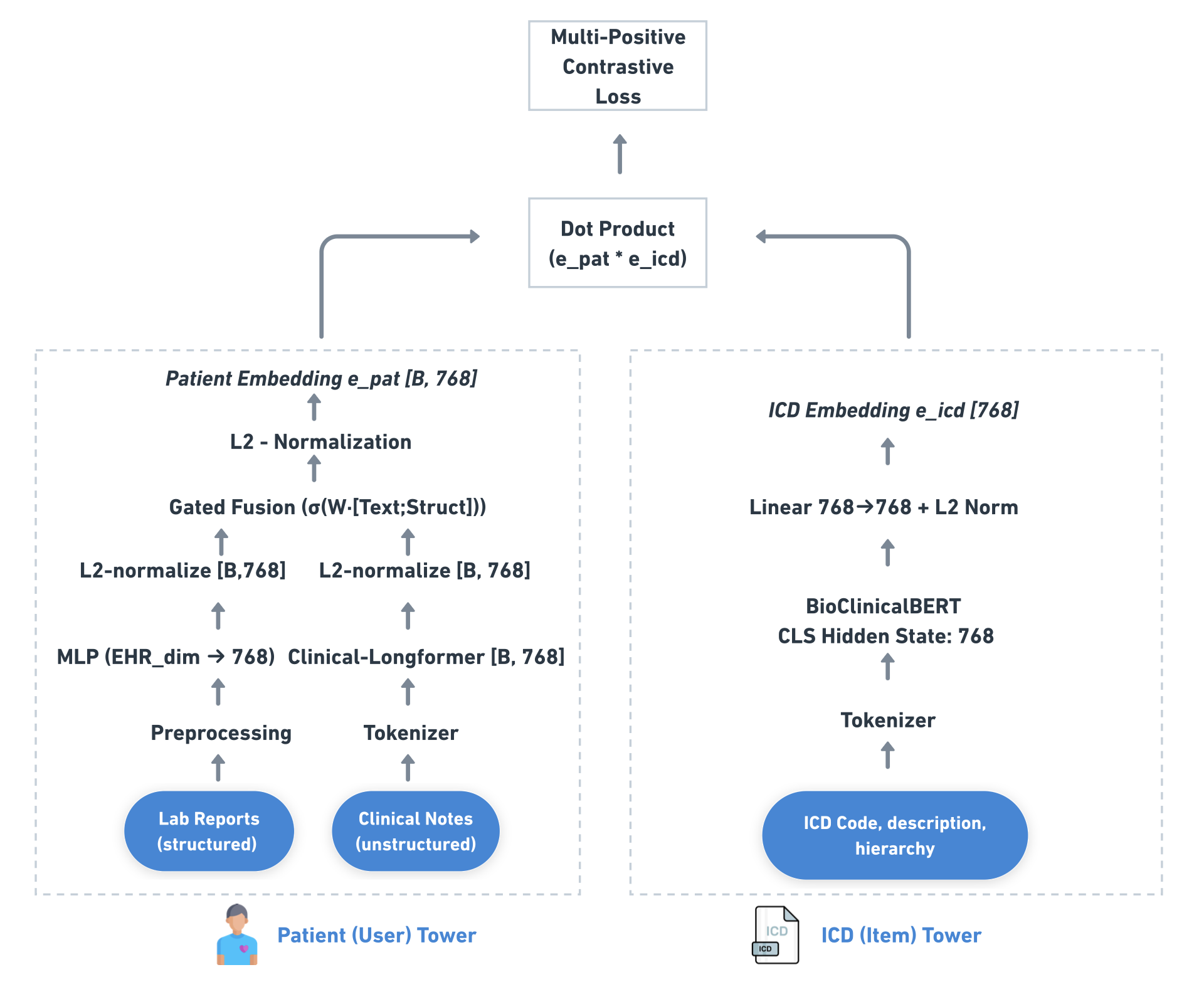}
    \caption{System Architecture (Training)}
    \label{fig:System_Training_Architecture}
\end{figure}

Fig.~\ref{fig:System_Training_Architecture} illustrates the training architecture of the proposed dual-encoder retrieval model. 

The patient (user) tower constructs a multimodal patient representation from two complementary sources: structured EHR features and unstructured clinical notes. The structured features are processed through a Multi-Layer Perceptron(MLP) and projected into 768-dimensional embedding space. In parallel, the unstructured clinical notes are encoded into a 768-dimensional embedding $z_{\text{text}}$ using Clinical-Longformer \cite{li2022clinicallongformerclinicalbigbirdtransformerslong}. Both modality-specific embeddings go through L2 normalization to stabilize training and ensure that dot product/cosine similarity corresponds to angular distance. The two normalized vectors are combined using a sigmoid gated fusion layer that learns per dimension mixture weights. This produces the final patient embedding $e_{\text{pat}}\in\mathbb{R}^{768}$.

The ICD (item) tower encodes ICD code metadata into the same 768-dimensional ICD embedding $e_{\text{icd}}$ using BioClinicalBERT \cite{alsentzer2019publiclyavailableclinicalbert} to enable similarity based retrieval. We extract the \texttt{[CLS]} token (hidden size 768) and pass it through a lightweight linear projection layer followed by L2 normalization. This gives the final code embedding $e_{\text{icd}}\in\mathbb{R}^{768}$ 

During training, both towers learn with a contrastive retrieval objective that allows clinically correct patient–code pairs to be embedded together, while incorrect pairs are pushed apart. Detailed components of this architecture are described in the following subsections.
\vspace{0.5mm}
\subsubsection{Patient Tower (User Tower)}
The Patient Tower generates a unified multimodal embedding by integrating information from two parallel sub-encoders: a text sub-encoder for unstructured clinical narratives and a structured sub-encoder for EHR data. A single encoder was avoided because it either forces numerical data to be serialized as weak tokens or causes text to dominate numeric features during optimization. Separate sub-encoders preserve modality specific biases and allow controlled fusion.

The text sub-encoder processes unstructured clinical narratives with a domain-specialized transformer (Clinical-Longformer). Clinical notes are long, so we segmented them into chunks, each chunk is encoded and mean-pooled. Then we aggregate chunk embeddings via attention to produce a single encounter vector $h_{\text{text}}$. The linear projection projects the encounter vector into the shared embedding space and then we apply normalization for stable training. 
\vspace{1mm}
\begin{equation}
z_{\text{text}}=\mathrm{LN}(W_t h_{\text{text}}+b_t)\in\mathbb{R}^{d}
\end{equation}
The structured sub-encoder is a lightweight feed-forward network designed to process numerical and categorical signals (e.g., lab data defined in the dataset section). This encoder is intentionally compact to keep the retrieval pipeline fast and stable across hospitals with heterogeneous lab panels.
\vspace{1mm}
\begin{equation}
z_{\text{struct}}=\mathrm{MLP}(x_{\text{struct}})\in\mathbb{R}^{d}
\end{equation}
Both modality specific embeddings are combined using gated fusion, which learns to adaptively weight text or structured data depending on clinical context, addressing the limitations of naive concatenation:

\begin{equation}
    \begin{gathered}
        u = [z_{\text{text}}; z_{\text{struct}}], \quad g = \sigma(W_g u + b_g) \\
        e_{\text{pat}} = g \odot z_{\text{text}} + (1-g) \odot z_{\text{struct}}
    \end{gathered}
\end{equation}

The final patient embedding $e_{\text{pat}}$ is followed by $l_2$ normalization. The resulting $e_{\text{pat}}\in\mathbb{R}^{d}$ is optimized for dual-encoder retrieval against ICD embeddings.
\vspace{0.5mm}
\subsubsection{ICD Code Tower (Item Tower)}
The ICD Code Tower encodes each ICD code into a dense semantic vector in the same vector space as the patient tower to enable scalable similarity based retrieval. Each ICD code is represented using available code metadata such as description, chapter/category and hierarchy path (parent -> child) to inject taxonomy structure. These features are provided to a biomedical text encoder (BioClinicalBERT) to generate a dense vector representation. Next, we apply a lightweight projection layer to align the embedding with the vector space followed by L2 normalization. The ICD tower is optimized jointly with the patient tower using contrastive retrieval defined in the next section, which aligns clinically correct patient-code pairs and separates mismatches in the same vector space. After training, ICD embeddings are computed and stored in a vector index for efficient ANN retrieval during inference.
\vspace{0.5mm}
\subsubsection{Dual-Encoder Contrastive Retrieval}
The Patient Tower and ICD Code Tower are jointly optimized using contrastive retrieval objective so that relevant patient-code pairs are embedded close in the shared semantic space. Given a normalized patient embedding $\hat e_{\text{pat},i}$ and ICD embedding $\hat e^{\mathrm{ICD}}_{j}$, patient–code affinity is computed using dot product. 
\vspace{1.5mm}
\begin{equation}
s_{i,j}=\langle \hat e_{\text{pat},i},\,\hat e^{\mathrm{ICD}}_{j}\rangle
\end{equation}

\textit{Multi-Positive Formulation}: Each patient visit may have multiple ICD codes assigned, therefore we use a multi-positive contrastive learning approach. For a given patient $i$, all ICD codes assigned to that visit form the positive set $P(i)$, while other codes in the batch serve as negatives. We adopted a multi-positive InfoNCE loss where $\tau$ is the temperature parameter and $\mathcal{B}_{\mathrm{ICD}}$ is the batch of patients in the current batch in the batch $\mathcal{B}$.

\begin{equation}
\mathcal{L}{\text{pat}\rightarrow\text{ICD}} = -\frac{1}{|\mathcal{B}|} \sum_{i\in\mathcal{B}} \log \frac{\sum_{p\in P(i)} \exp\!\big(s_{i,p}/\tau\big)} {\sum_{j\in \mathcal{B}_{\mathrm{ICD}}} \exp\!\big(s_{i,j}/\tau\big)}
\end{equation}

Negative samples were generated using two approaches: (i) In-batch negatives: Every non-positive code present in the current batch acts as an implicit negative for patient i. (ii) Hard negatives: We sample codes that are clinically similar to the positives to sharpen fine grained distinctions. Batches are built to maximize in-batch negatives and include hard negatives from the same ICD chapter when available to ensure semantic proximity. We use standard weight decay and compute the softmax in log-space for numerical stability. This training approach enables appropriate ICD codes rank highly for each patient. The top-k ICD codes are passed to the downstream re-ranking component for further refinement.  

\subsection{Inference Architecture}

\begin{figure}
    \centering
    \includegraphics[width=1.07\linewidth]{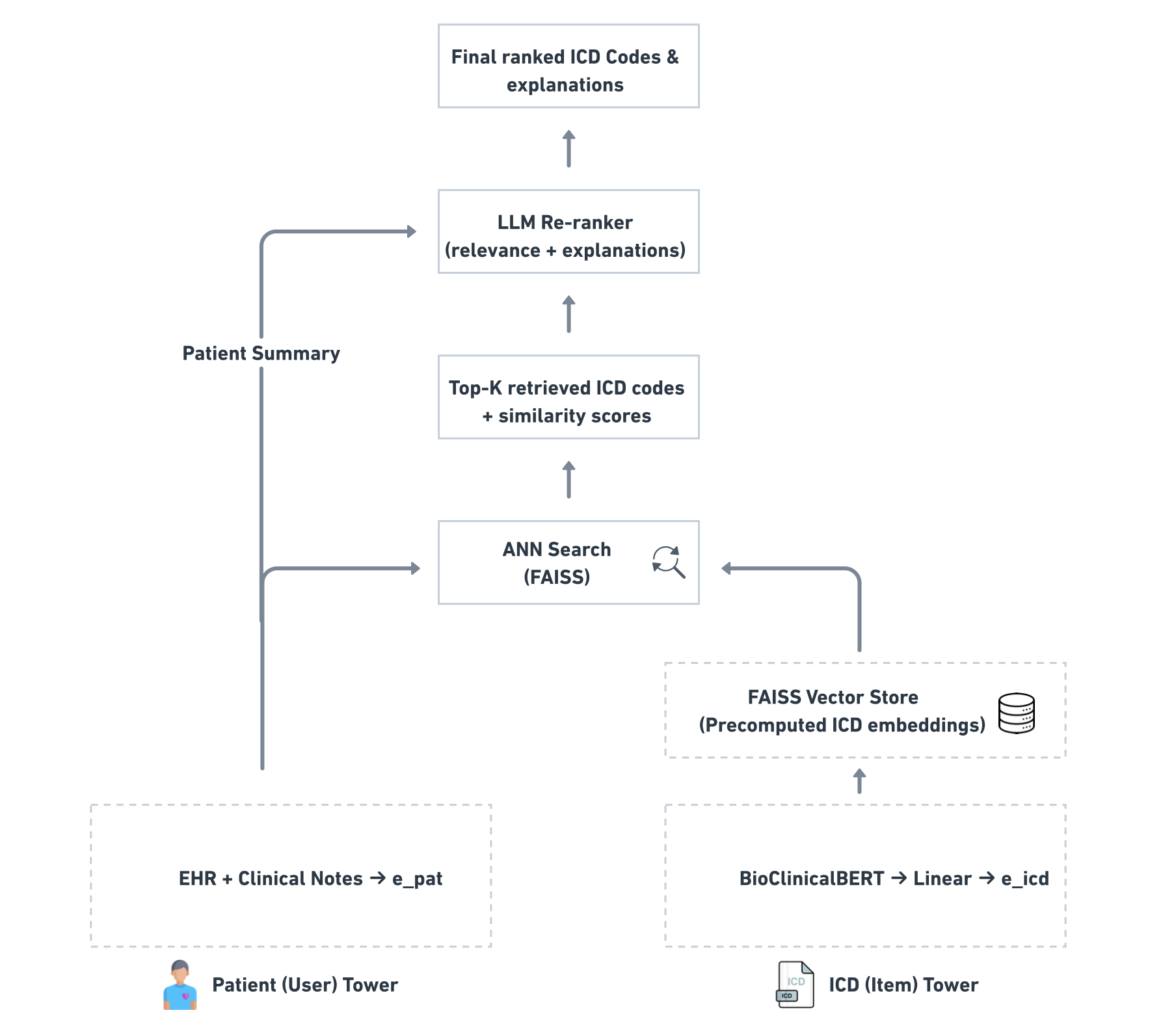}
    \caption{System Architecture (Inference)}
    \label{fig:System_Architecture_Inference}
\end{figure}

Fig.~\ref{fig:System_Architecture_Inference} illustrates the inference architecture. After training, we freeze the ICD code tower and run an offline job over the full ICD catalog to generate dense embedding for each ICD code. Each embedding is L2-normalized and stored in a parquet embedding store. 
We then build an ANN index using Facebook AI similarity search (FAISS) \cite{johnson2017faiss}. The FAISS index is stored as a serving artifact and it can be reloaded without recomputing code embeddings.

During inference, for each patient admission, we run the patient tower to generate a normalized patient embedding. After that, the embedding is used to query the FAISS index to obtain top-$K$ codes and their similar scores. Finally, the retrieved code list is passed to the LLM re-ranking module for further refinement. 

\subsection{LLM Re-ranking}
The LLM-based re-ranking component reorders the top-$K$ ICD codes returned by the retriever model and simultaneously provides human-readable explanation. While the retriever efficiently generates a candidate set, it may assign similar scores to semantically close but clinically different codes. To resolve this, we use a second stage to reason over the patient and ICD code context to make finer distinctions. We use claude 3.7 sonnet \cite{anthropic2025claude37} in a few-shot setting to reassess the clinical relevance of each candidate. For a given patient $i$, we first construct a textual summary $p_i$ that concatenates important elements from structured EHR features (e.g., key lab values, demographics) and unstructured clinical notes (e.g., symptom descriptions, treatments). 

We then present $p_i$ together with candidate ICD code metadata $c_j$ in a natural language prompt, the prompt includes a brief patient summary, the ICD code description, and available category or hierarchy information. We ask the LLM to judge how well the code matches the case and to briefly justify the decision. The LLM outputs a JSON output, for each patient-code pair, it provides a scalar relevance score $r_{i,j}$ and a short explanation $e_{i,j}$.
\vspace{1.5mm}
\begin{equation}
\{(r_{i,j}, e_{i,j})\}_{j=1}^{k} \;=\; f_{\mathrm{LLM}}\!\left(p_i, \{c_j\}_{j=1}^{k}\right),
\end{equation}
where $f_{\mathrm{LLM}}(\cdot)$ denotes the LLM’s relevance estimate induced by the prompt. In practice, $r_{i,j}$ can be derived from the LLM’s ranking of candidates or from explicit relevance scores returned in the JSON output. These scores are used to reorder the top-$K$ codes produced by the retriever. In explanations, the LLM highlights key phrases or findings in the patient summary that support its decision. For example, it may output: \textit{``Persistent high fasting glucose and ongoing insulin therapy strongly indicate type 2 diabetes with hyperglycemia''}. Such explanations support clinical review and improve trust.

\section{Experiments}
\subsection{Dataset}
We evaluate our approach using the MIMIC-III v1.4 critical care dataset \cite{johnson2016mimic}, which contains EHRs of 53,423 ICU related hospital admissions from 2001-2012. Each sample corresponds to a single hospital admission with ICD-9 codes. To avoid leakage across multiple records for the same patient, the dataset was split at patient level into 70\% training, 15\% validation and 15\% test sets. The dataset consists of structured laboratory and demographic data, and unstructured clinical notes for the patient tower. Structured EHR features included laboratory reports containing lab test names, values, admission metadata, along with demographics, admission metadata (admission type, ICU unit). The unstructured data included clinical notes, discharge summaries, radiology reports, and physician notes. The ICD tower uses code metadata: code identifier, short description, chapter information/ICD category, and a hierarchy path that outlines the parent–child structure of the code within the ICD classification system (e.g., “Endocrine disorders → Diabetes mellitus → Type 2”)

\subsection{Evaluation}
Each patient admission from the MIMIC-III dataset has a set of ICD-9 codes without an inherent ranking or score. The proposed retrieval model outputs a ranked list of ICD codes with similarity score for each admission. Therefore, we evaluated performance using metrics that capture both retrieval and ranking quality. Across experiments, we report Precision@K (where k $ \in $ 3, 5, 8) on the retrieved ranked list.  Additionally, we report Micro-F1@K score using top-k decoding rule for the classification style comparison across methods, where the top-k retrieved ICD codes are treated as positives. For classifier baselines, ICD codes are ranked by predicted probability and top-k codes are used to calculate Precision@k and Micro-F1@k The LLM-based re-ranking is evaluated using NDCG@8 on a human annotated subset with graded relevance labels provided by clinical annotators, since the MIMIC-III does not provide ground truth rankings. Human labels are collected using a 3 level rubric where 2 is correct match, 1 is partially related, 0 is irrelevant. Two humans labelled a stratified subset of N=2000 samples to measure agreement, and remaining cases are single annotated.

\subsection{Experimental Setup}
This section defines the baseline models to benchmark the performance of the proposed model. We trained and evaluated performance on the same dataset. The initial baselines create separate models for structured and unstructured data.

{1. XgBoost:} 
The XGBoost model is trained using structured EHR features, which include numerical laboratory values and demographic features. We treated ICD prediction as a multi-label classification task by using the XGBoost V2.0 \cite{chen2016xgboost} vector leaf tree models. Unlike one-vs-rest, where it trains one model per ICD code, vector leaf trees learn a single ensemble in which each tree leaf stores a vector of scores. It outputs one confidence score per ICD code from a single forward pass. This is implemented by adding these parameters: \texttt{multi\_strategy="multi\_output\_tree"} and \texttt{tree\_method="hist"}. We even tuned parameters such as learning rate, tree depth, and number of estimators. The model captures non-linear interactions among structured variables through gradient-boosted decision trees, making it suitable for tabular medical data. This baseline accesses predictive power in tabular structured data alone, without any unstructured clinical notes.

{2. BioClinicalBERT:}  BioClinicalBERT is a transformer-based language model pretrained on large-scale biomedical and clinical texts. In this baseline, we used unstructured clinical notes textual data as input. We tokenized and concatenated the discharge summaries, medications, previous prescriptions and symptom descriptions using BioClinicalBERT, followed by a fully connected multi-label classification head applied to the contextual embeddings. The model is fine-tuned using binary cross-entropy loss over all ICD codes, with a learning rate of $2e^{-5}$, Adam optimizer, and early stopping based on validation Micro-F1. This baseline extracts semantic features from clinical text but had limited information about structured lab and demographic features.

{3. Dual Fusion Encoder:}
 The dual fusion model is multimodal, it consisted of two separate encoders, one for structured data encoder which processed lab reports and demographic information using a feed forward neural network with multiple hidden layers and ReLU activation function. The other unstructured data encoder was a clinical-longformer model to extract generate embeddings for unstructured features. The output of the structured and unstructured encoders are concatenated and passed through a fusion layer followed by a multi-label classification head. The model is trained using binary cross-entropy loss function, using Adam Optimizer and early stopping based on validation Micro-F1. This baseline leverages both modalities, but remains a conventional classifier that produces logits for each ICD code with limited scalability. This led to the inspiration of creating a dual-encoder retrieval system.

 {4. Proposed Dual-Encoder Retriever + LLM Re-ranker:}
Our proposed framework performs ICD code retrieval followed by LLM re-ranking. The patient tower outputs multimodal embedding by fusing two sub-encoders: clinical-longformer for unstructured clinical notes and MLP encoding structured EHR features. Both modality vectors are projected into a shared 768 dimensional normalized embedding space. The ICD tower encodes code metadata using BioClinicalBERT. We train both towers jointly using multi-positive InfoNCE loss function with AdamW optimizer and a batch size of 256. Each admission can have multiple positive ICD code labels, so we constructed negatives using in-batch negatives and hard negatives (sampled from same ICD chapter when available). We then freeze the ICD tower to precompute embeddings and store them in a parquet backed embedding table. During inference, we build a FAISS ANN index and retrieve top-k candidates. For re-ranking, we use the claude 3.7 sonnet model \cite{anthropic2024claude} to assign a relevance score for each patient-code pair and evaluate the ranking quality using NDCG@k on 2000 human annotated subset.
\vspace{-1mm}
\section{Results}
\subsection{Baselines vs Retriever model on MIMIC-III} 
\begin{table}[h]
\centering
\caption{Performance comparison between ICD coding baseline models and the proposed dual-encoder retriever.}
\label{tab:baseline_comparison}
\setlength{\tabcolsep}{10pt}
\resizebox{\columnwidth}{!}{
\begin{tabularx}{\columnwidth}{lcccc}
\toprule
\textbf{Model} & \textbf{P@3} & \textbf{P@5} & \textbf{P@8} & \textbf{Micro-F1@8} \\
\midrule
XgBoost & 51.43 & 48.32 & 42.34 & 52.11 \\
BioClinicalBERT & 56.10 & 53.38 &  45.91 & 53.51 \\
Dual Fusion & 62.12 & 58.92 & 51.38 & 59.73 \\
Retriever Model & 66.75 & 62.66 & 54.23 & 62.44 \\
\bottomrule
\end{tabularx}}
\end{table}

Table \ref{tab:baseline_comparison} presents an evaluation comparing the retrieval dual-encoder model to the baseline models. Each model was trained and tested on the same dataset to ensure consistency in benchmarking. Overall, our proposed retrieval model outperforms all three baselines across all measured metrics. XGBoost attains a reasonable Micro-F1 score, which indicates that structured features carry considerable predictive signal. Next, BioClinicalBERT further improves the performance metrics, demonstrating the importance of unstructured clinical narratives for ICD code prediction. Then, the Dual Fusion improves over both single modality baselines in Table 1. By jointly encoding structured and unstructured input, it clearly benefits and achieves higher Micro-F1. Finally, the proposed dual-encoder retrieval model shows a notable gain in Precision@3 from 62.12\% to 66.75\% when compared to the strongest baseline (Dual Fusion). These gains indicate that modeling ICD coding as a retrieval problem yields high quality output than multi-label classification. We analyzed the impact of LLM based re-ranking separately in the next section.

\subsection{LLM-Re-Ranker and Explainability}
The top-k ICD codes produced by the retrieval model are further refined using an LLM layer. We employ a Claude 3.7 Sonnet model to assess relevance between each patient-code pair. The LLM takes unstructured clinical notes and structured features along with ICD code metadata as input and outputs a relevance score and a short justification for each patient-code pair, which are used to reorder the top-k ICD codes. Because MIMIC-III does not provide ranked labels or score, we evaluated the LLM re-ranked codes using human in the loop. The human annotated dataset consisted of 2000 randomly sampled patients from the MIMIC-III test set. The ranking quality was measured using NDCG@8, with LLM re-ranking NDCG@8 improved from 65\% to 71\%.


\section{Conclusion}
We present a two stage system that integrates dual-encoder retrieval which combines multimodal patient data with ICD code metadata and LLM-based re-ranking with reasoning capabilities. Across MIMIC-III, the proposed approach improves Precision@k and Micro-F1 over multimodal baseline, and LLM re-ranking further improves the NDCG@k on a human annotated dataset. While the LLM re-ranking component showed gains in ranking quality, we need further large scale clinical validation. These results suggest that retrieval based approach offers a scalable alternative to dense multi-label classification. Further validation is needed before deployment in real clinical coding workflows.

\section{Limitations and Future Work} 
Although the retriever system improved performance on the MIMIC-III dataset, the framework is limited to a single center ICU population and has yet to be validated against larger ICD-10 label space. Additionally, while the LLM improved accuracy, its high computation cost, additional latency, and potential for clinical hallucination present challenges for real-time deployment. Therefore, it should be treated as an optional second stage for ambiguous cases rather than as a replacement for the low-latency retriever. Future work will focus on scaling the system by utilizing knowledge distillation to create efficient student re-ranker, and doing rigorous clinical validation to ensure faithfulness of the generated explanation.

\def\IEEEbibitemsep{0pt plus .5pt}
\begingroup
\footnotesize
\bibliographystyle{IEEEtran}
\bibliography{reference}
\endgroup

\end{document}